\documentclass[manuscript]{acmart} 

\usepackage{subcaption}
\usepackage{tabularx}

\AtBeginDocument{%
  }

\setcopyright{acmlicensed}
\copyrightyear{2018}
\acmYear{2018}
\acmDOI{XXXXXXX.XXXXXXX}
\acmConference[Conference acronym 'XX]{Make sure to enter the correct
  conference title from your rights confirmation email}{June 03--05,
  2018}{Woodstock, NY}
\acmISBN{978-1-4503-XXXX-X/2018/06}

\begin{document}

\title{Patterns in the Transition From Founder-Leadership to Community Governance of Open Source}

\author{Mobina Noori}
\affiliation{%
  \institution{University of California Davis}
  \department{Department of Computer Science}
  \city{Davis}
  \state{CA}
  \country{USA}
}
\email{lianoori@ucdavis.edu}

\author{Mahasweta Chakraborti}
\affiliation{%
  \institution{University of California Davis}
  \department{Department of Communication}
  \city{Davis}
  \state{CA}
  \country{USA}}
\email{mchakraborti@ucdavis.edu}

\author{Amy X. Zhang}
\affiliation{%
  \institution{University of Washington}
  \department{Allen School of Computer Science \& Engineering}
  \city{Seattle}
  \state{WA}
  \country{USA}
}
\email{axz@cs.uw.edu}

\author{Seth Frey}
\affiliation{%
  \institution{University of California Davis}
  \department{Department of Communication}
  \city{Davis}
  \state{CA}
  \country{USA}}
\email{sethfrey@ucdavis.edu}

\renewcommand{\shortauthors}{Trovato et al.}

\begin{abstract}

Open digital public infrastructure needs community management to ensure accountability, sustainability, and robustness. Yet open-source projects often rely on centralized decision-making, and the determinants of successful community management remain unclear. We analyze 637 GitHub repositories to trace transitions from founder-led to shared governance. Specifically, we document trajectories to community governance by extracting institutional roles, actions, and deontic cues from version-controlled project constitutions (\texttt{GOVERNANCE.md}). With a semantic parsing pipeline, we cluster elements into broader role and action types. We find roles and actions grow, and regulation becomes more balanced, reflecting increases in governance scope and differentiation over time. Rather than shifting tone, communities grow by layering and refining responsibilities. As transitions to community management mature, projects increasingly regulate ecosystem-level relationships and add definition to project oversight roles. Overall, this work offers a scalable pipeline for tracking the growth and development of community governance regimes from open-source software's familiar default of founder-ownership.
\end{abstract}





\maketitle

\section{Introduction}
Open source software (OSS) underlies critical infrastructure around the globe, and has a hand in every aspect of digital life. And yet, most OSS projects are run as “founder-led, single-leader governance models” with no formal accountability to their community, or society.
OSS projects are largely volunteer-driven initiatives in which developers join, cooperate or cease contributing of their own volition. Although the top 1000 projects on the popular OSS hosting site GitHub show an average of ~80 contributors \cite{bao2021large}, projects on such platforms start with just one founding contributor, by design, and most remain in a regime under the monarchic default \cite{schweik_internet_2012}. It is on founders to define governance processes: the specifics of contributing, or making and pursuing fixes and features, determining the rights of financial supporters, and defining the larger goals of the project \cite{o2007emergence}. This implicit single-leader model is likely appropriate for the early stages of a project \cite{schneider_modular_2021}. But as a project becomes important to its contributors, and even society, a founder-owner model becomes increasingly inappropriate from an accountability standpoint, particularly when a piece of software becomes critical digital public infrastructure \cite{zuckerman2020case}. Although many OSS ventures begin with founder-centered, centralized control, sustaining growth and broad participation typically requires evolving governance to include shared leadership and community norms, thereby distributing influence beyond a single individual or core team \citep{10.1145/1061874.1061885}\citep{10.1145/3274326}.

\subsection{Transitions to community governance in open-source software (OSS)}
Fortunately, OSS projects seem to agree with this assessment. It is now commonplace that the Internet’s most prominent OSS projects are undergoing transformations in their governance from the founder-owner default to community governance. After six years of operating under an ideologically aligned founder, but without formal community representation, the Debian operating system in 1997 initiated a multi-year governance transition process that concluded in annual democratic elections (which, themselves, continued to evolve in form and scope for several years) \cite{o2007emergence}.  The Django web framework moved from a single-leader governance model to a core team  model in 2014 \cite{holovaty2014django} and from a core team model to a more inclusive and community-driven “technical advisory board” in 2020 \cite{bennett2020django}. The developers of the Python programming language, after decades under a single-leader governance model, took inspiration from the Django transition \cite{warsaw2018python} to implement an elected board structure.  In social media, the growth of the early community-led news website Slashdot motivated several redesigns of the site’s moderation system, from manual moderation by the founder to ever-expanding methods of delegating the task to users, ultimately giving reputable users considerable opportunities for effective voice over Slashdot’s content \cite{ganley2009slashdot}.  Other prominent examples of OSS projects that have transitioned from founder-owners to community governance include the FreeBSD operating system and the Apache server. At the same time, notable projects such as the Linux operating system and Vim text editor have persisted under founder-owner leadership.

However encouraging and important, the phenomenon of voluntary transitions to community is poorly understood. Each case has enough important differences to obscure any core similarities. They may initiate their transitions voluntarily or involuntarily, because of core team burnout, community demands, or even the death of a founder, as in the 2017 case of OpenDataKit. During transitions, they use different tools to help them through the process, often adapting the tools they have for the needs they discover. Python, for example, solicited governance proposals from its community through its preexisting system for evaluating technical language improvements, the Python Enhancement Proposal (PEP) system \cite{edge2018python}. While reflecting the strength and determination of OSS communities to steer themselves toward accountability \cite{izquierdo_role_2018}, such governance software “hacks” reflect the OSS community’s need for shared, general, theoretically-grounded tooling dedicated to the challenging problem of major governance transformation. Overall, the trend is happening in large part in spite of available models and tools, rather than because of them. Legal, technical, and even cultural defaults lead the overwhelming majority of OSS projects start under founder-led, single-leader governance, and only much later, if at all, transfer ownership and management (and thus, accountability) to their communities \cite{schneider_modular_2021}.

Community governance of OSS not only secures the accountability of a project’s design decisions to society, but may also reduce the administrative load on project founders, and subsequent burnout \cite{hertel_motivation_2003, 10.1016/j.infoecopol.2008.05.001}. However, the governance defaults of open source software hosting platforms do little to ensure the success of community management \cite{schneider_modular_2021}. As the stakes of a project increase, so do its responsibilities to its community, particularly if that project finds itself playing a role in critical public infrastructure (Goggins et al., 2021; Overney, 2020). Thus, the absence of robust support for community management can plant the seeds for society-scale accountability problems down the line, as OSS projects on those platforms become more successful.

\subsection{Governance documents as a window into institutional processes}
Prior research has tended to approach OSS governance through indirect signals and descriptive accounts rather than through systematic, longitudinal documentation of governance as expressed in textual artifacts. Empirical work has largely inferred governance from behavioral traces (such as commits, issues, and mailing lists) and complemented this with qualitative case studies that detail governance mechanisms, including contribution guidelines and social practices that support coordination in distributed communities \citep{10.1145/1882362.1882446}\citep{10.5465/amj.2007.27169153}. Much of the prior work tends to focus on more superficial features of policy texts, such as their length, headers, and topics \cite{beck_complexity_2003}, and to be cross-sectional rather than tracking persistent governance inscriptions over time \citep{10.1145/1061874.1061885}\citep{10.1145/3570635}.

Thus a significant gap remains: there is no established longitudinal account of how governance transitions to community management are executed in formal textual artifacts and how these representations evolve during this institutional interregnum. This gap motivates a textual-governance lens in OSS governance evolution, acknowledging that governance-related content encoded in artifacts—such as guidelines, policies, and governance-oriented files stored in version control—provides direct insight into institutional structure and accountability, in the process of aligning with broader calls for public stewardship and accountability in governance of open source software \citep{10.1145/2531602.2531659}\citep{10.1145/3570635}\cite{zuckerman2020case}. 

\subsection{Project governance on GitHub: The \texttt{GOVERNANCE.md} standard}

GitHub is the most widely used hosting platform for open-source development, built on the distributed version-control system Git. It provides an infrastructure for collaboration, coordination, and community visibility as well as storing code. Governance is a persistent concern in this context: projects must determine how authority should be allocated, how contributor rights should be granted or lost, and how conflicts should be resolved. 

Fortunately, as a reflection of its resourcefulness, the OSS community has developed a loose standard for defining OSS project governance on GitHub, one that prescribes a clear, uniform location and format for governing documents. Increasingly, influential OSS projects on GitHub—such as node.js, Docker, and Jupyter—are posting public \texttt{GOVERNANCE.md} files that list out their policies for contributions, donations, planning, and other essential OSS governance functions. \texttt{GOVERNANCE.md} files  are an elegant use of the “git” version control software to assign responsibilities and provide clear community entry points into OSS governance \cite{izquierdo2015governance}. In this it follows other emergent conventions for community management, like \texttt{CONTRIBUTING.md} and \texttt{CODE\_OF\_CONDUCT.md} files. Also, because \texttt{GOVERNANCE.md} is a plain text file co-hosted with project code, it can leverage git’s tools for version control and forking. This means tracking of historical changes by default, and excellent tools for projects to learn from each other’s governance designs, and select their own institutional designs in an informed manner.
\texttt{GOVERNANCE.md} files are an entry point for OSS projects to transition from the implicit feudalism of founder-owners to true, and truly accountable community governance. 

\subsection{The study}
To address the gap in our understanding of governance transitions from unitary to community governance, we compile a dataset of 637 open-source repositories to compare the initial and latest snapshots of the \texttt{GOVERNANCE.md} constitutions, enabling a longitudinal examination of how governance is explicitly articulated and evolves across projects \citep{yin_open_2022}\citep{10.1109/icse.2015.184}.We parse text from these artifacts into three core components—Roles, Actions, and Deontics—capturing who is responsible, what activities are authorized or required, and the normative force behind those directives, reflecting established practices for articulating governance rules and governance-oriented roles in OSS ecosystems \citep{10.1109/icse.2015.184}\citep{10.1007/s10664-021-10061-x}. The methodological contribution lies in integrating NLP-driven information extraction with statistical approaches to scale OSS governance analysis across hundreds of projects: we leverage textual embeddings and grouping techniques to surface governance patterns from many repositories, a direction supported by prior work that combines linguistic analyses of OSS communications with longitudinal network perspectives to study governance dynamics \citep{yin_open_2022}\citep{10.1109/icse.2015.184}.This approach operationalizes governance as a textual, machine-readable artifact and enables cross-project comparability, consistent with the literature's emphasis on explicit governance documentation and the use of computational methods to analyze governance-related text in relevant contexts \citep{10.1109/icse.2015.184}\citep{yin_open_2022}.

Building on this dataset and analytic pipeline, we capture governance evolution with two measures of the institutional reflected in \texttt{GOVERNANCE.md} snapshots. Entropy ($H$) captures the evenness of the distribution of governance categories (e.g., roles, actions, deontics) across a snapshot; higher $H$ indicates a more uniform allocation of governance attention across categories, while lower $H$ signals concentration in a subset of categories, a concept rooted in entropy estimation and its application to software governance data \citet{10.1007/978-3-031-21388-5_20}. Count ($K$) reflects the diversity of governance elements present at a given version, namely the number of distinct roles, actions, and deontics formalized, drawing on OSS diversity research that links governance variety to ecosystem outcomes and sustainability \citep{10.1145/2961111.2962633}\citep{10.1287/isre.1120.0435}. 
Together, $H$ and $K$ provide a compact, quantitative  characterization of institutional development over time, enabling cross-project longitudinal comparisons and informing understanding of how governance structures emerge and evolve in distributed OSS ecosystems \citep{10.1145/2961111.2962633}\citep{10.1287/isre.1120.0435}.

Together, this dataset, analytic pipeline, and measurement framework underpin four key contributions of this work. This work provides (i) a dataset comprising paired \textit{initial} and \textit{latest} \texttt{GOVERNANCE.md} snapshots from 637 OSS projects, enabling longitudinal study of explicit governance articulations across diverse ecosystems \citep{10.15209/jbsge.v6i1.195}\citep{10.1007/s10997-007-9021-x}; (ii) a scalable NLP pipeline for extracting institutional components (Roles, Actions, Deontics) from governance text, surfacing formal structures across repositories \citep{10.15209/jbsge.v6i1.195}\citep{10.1007/s10997-007-9021-x}; (iii) a measurement framework employing Entropy ($H$) and Count ($K$) to quantify governance change over time, integrating insights about governance scope and distributional changes \citep{10.1007/978-3-031-21388-5_20}\citep{10.1145/2961111.2962633}\citep{10.1287/isre.1120.0435}\citep{10.48550/arxiv.0804.1653}\citep{10.14736/kyb-2021-6-0879}\citep{10.3390/e21050485}; and (iv) empirical findings demonstrating that OSS governance diversifies and rebalances as projects evolve, signaling maturation of governance structures beyond initial configurations \citep{10.2139/ssrn.474782}\citep{KroghSpaeth2003}\citep{10.5033/jolts.v11i1.131}. Together  these contributions illustrate how digital communities formalize and transform governance, offering both theoretical and methodological contributions to collaboration and information-systems research by foregrounding textual governance as a deployable, comparable artifact, and by providing scalable tools and metrics for population-level analyses \citep{10.15209/jbsge.v6i1.195}\citep{10.1007/s10997-007-9021-x}\citep{Hill_Shaw_2017}.

\section{Related Work}

Governance in open-source software is a fundamental determinant for sustaining critical digital infrastructure, as the organization of authority, responsibility, and decision-making directly shapes participation, accountability, and conflict resolution across distributed communities \citep{10.1145/1061874.1061885,sundaramurthy2003control}.

Importantly, governance encompasses more than technical coordination; it is a social process that structures participation, trust, and accountability in diverse, geographically dispersed collaboration networks, with group awareness, norms, and transparent decision-making playing central roles in sustaining coordination \citep{10.1145/1031607.1031621}. Empirical analyses across OSS projects emphasize that governance mechanisms—such as conflict management, contribution guidelines, and social practices that facilitate knowledge sharing—are essential to long-term resilience, not merely auxiliary to code development \citep{10.1145/1061874.1061885}\citep{10.1145/2531602.2531659}\citep{10.1145/2089125.2089127}. Consequently, OSS governance emerges as a socio-technical concern that not only warrants scholarly attention but also raises the need for approaches that can directly capture how governance is articulated and evolves within projects \citep{noori2025human}\citep{10.1145/3274326}\citep{10.1145/1031607.1031621}.

Research on governance in open-source software (OSS) has examined a wide range of mechanisms through which communities establish norms, coordinate work, and manage participation. Early studies highlight the role of policy instruments such as Codes of Conduct (CoCs), which are frequently deployed to set norms and manage behavior within OSS communities, functioning as governance mechanisms that constrain harassment and define participation rules. Their adoption and negotiation are inherently political and organizational processes that shape community identity and power distribution \citep{10.1145/3449093}. Complementing these community-level instruments, Harutyunyan and Riehle \citep{harutyunyan_getting_2021} study how organizations begin implementing OSS governance, offering case-based insights into corporate entry points and best practices. Beyond formal artifacts, Gibbs et al. \citep{10.1093/ct/qtab017} theorize how concertive control and coordination processes emerge in online communities, showing how norms evolve in practice through communicative processes such as Linux kernel discussions. Similarly, Drost-Fromm and Tompkins \citep{10.1109/mc.2021.3058023} emphasize the human and community aspects of governance, using the Apache Way as a case study to articulate how people and relationships remain central to sustaining OSS projects.

Building on these perspectives, subsequent work has examined governance at the level of contributions and technical-social decision-making. In accounting for how contribution management blends technical evaluation with social considerations, Alami et al. \citep{10.1109/tse.2021.3128356} identify three governance styles in the handling of pull requests: protective, equitable, and lenient. In parallel, Ghag \citep{10.47363/jaicc/2022(1)249} develops a theoretical framework linking community engagement models, governance structures, and OSS security practices, underscoring governance as a driver of both participation and system-level resilience.

\section{Theoretical Background}
\subsection{Institutional analysis and development}
Institutional analysis and development (IAD) is a framework for analyzing self-governance in small-scale communities \cite{ostrom2005understanding}. It was developed by the research community that developed around political scientists Vincent and Elinor Ostrom to attend to the tragedy of the commons in natural resource management systems around the world \cite{ostrom2010beyond}. Through their work, spanning over 50 years, they succeeded in reversing the tide of skepticism that small communities can overcome confounding collective action problems like the tragedy of the commons \cite{hardin1968tragedy}, and often outperform both state and private actors \cite{ostrom2006polycentric}. 

A major interest of the IAD community is in how community management of natural resources, as opposed to either private or state management, leverages its accountability to a wide range of stakeholders to evolve in a way that is sensitive to local resource conditions, and local community needs \cite{ostrom1990governing}. For this reason, IAD is an ideal starting point for an investigation of community governance in OSS projects. Within this framework, OSS projects, like other natural resource management communities, develop complex institutional structures to craft systems of incentives that secure their resilience to resource collapse.

Particularly useful are IAD’s tools for analyzing written policy texts. The institutional grammar is a linguistic approach to policy analysis developed by Sue Crawford and Elinor Ostrom for formally representing resource management institutions from their written or spoken rules \cite{crawford_grammar_1995, siddiki_assessing_2014}. Its basic unit is the “institutional statement” (IS), usually a sentence of policy text, which can be a rule, norm, or even a piece of institutionally relevant advice. An IS generalizes the idea of rule: it can be written or not, normative or not, and enforceable or not. When decomposed, institutional statements provide the basic elements of governance: the types of resources and agents that must be managed, as well as the conditions and nuances under which that management is more or less strict. It accomplishes this by mapping these institutional constructs—actors, actions, constraints, conditions, and other basic elements—to the linguistic elements of grammar and syntax \cite{crawford_grammar_1995, frantz_institutional_2021}. This work's extraction of project roles, actions, and deontics is drawn directly from Ostrom's institutional grammar

A growing body of literature is attending to online community governance, particularly under the lens of IAD. In digital contexts, the work of Ostrom is most closely associated with participatory self-governing online institutions \cite{schweik_internet_2012, hess2007understanding, Grudin1994ComputersupportedCW, Kittur2013FutureCrowdWork, Salehi2015WeAreDynamo, Silberman2016ReadingOstrom}. Ever since 1996, with Charlotte Hess's analysis of the Internet as a commons~\cite{hess1996untangling}, and Peter Kollock's similarly Ostromian analysis of USENET \cite{Kollock1996ManagingVirtualCommons}, the resource management perspective has had a small foothold in the study of digital institutions.  Major contributions have been volumes such as “Knowledge Commons” \cite{hess2007understanding}, “Internet Success” \cite{schweik_internet_2012}, and “Building Successful Online Communities” \citep{kraut2012building}. The institutional analysis lens has proven especially fruitful for understanding the community managed encyclopedia Wikipedia  \cite{forte2009decentralization, heaberlin_evolution_2016, viegas2007talk} (which underwent its own transition to community management from benevolent dictator Jimmy Wales). 

\subsection{Institutional development of online communities}

As this body of work advances, there is growing attention to community change processes \cite{centivany2016popcorn, heaberlin_evolution_2016, keegan2015is}, including full case analyses of prominent OSS governance transitions from founders to their communities \cite{o2007emergence, omahony2022proprietary, 10.1016/j.infoecopol.2008.05.001, yin_sustainability_2021, yin_open_2022}. Scholars have yet to extend these analyses beyond individual cases, or ground them systematically into existing constructs for conceptualizing institutional evolution.

If the status quo is any indication, online communities such as OSS projects will require active support for transitions to community ownership to become the norm \cite{schneider_modular_2021}. In work on a large ecosystem of self-hosted game communities based on the implicit feudalism default, Frey and others found that the most successful communities were those that empowered their administrators even more \cite{frey_emergence_2019, muller2015heapcraft}. Conversely, another analysis, of Wikipedia’s open, participatory, community-driven policy development process, finds that one third of proposals are abandoned, suggesting its failures to fully engage community members in its overtly community-centered governance processes \cite{im_deliberation_2018}.

\subsection{Policy perspectives on OSS self-governance}

The challenge of designing accountable software systems is easily construed through the lens of government policy and regulation. By what regulatory mechanisms and with what technical support can critical web technologies become accountable to the public \cite{mannan2021platform}. However, there is a more direct accountability, one with a strong legal track record, and increasing influence in the web’s digital public infrastructure: community ownership. Community ownership shortcuts large-scale regulatory schemes by making organizations directly accountable to their communities \cite{arnstein2007ladder}. Legal scholar Paul Gowder argues that centuries of political theory on states suggest the use of democratic institutions in platforms \cite{gowder2023networked}. Peer production scholar Siobhan O’Mahoney connects community governance to software system accountability by identifying five properties that community governance brings to OSS projects: independence, pluralism, representation, decentralized decision-making, and autonomous participation \cite{o2007emergence}.  

\subsection{Design perspectives on OSS self-governance}
Software tools for supporting community governance include tools for code review, code sharing, and the delegation of power and responsibility. Still, these tend to be built for single leaders and the users they entrust with power. Take the example of Reddit's Automod~\cite{jhaver2019human}, which gave subreddit moderators on the platform a rich language for handling tedious moderation tasks. As much as it can be appreciated for empowering the volunteer community leaders who rely on it, it is a clear example of a tool premised on assumptions of strong centralized community leadership. 

Fortunately, alongside the ever-growing suite of moderator support tools there is a parallel history of governance tools that keep leadership accountable to member needs (Ludlow, 2001). These range from the VOTEMGR software available for the early FidoNet bulletin board system (Castillo, 1991) to Loomio (Jackson \& Kuehn, 2016), which provides a platform for deliberation and community building as well as various voting mechanisms.  These are being followed with a new generation of self-governance tools that can be tailored to fit the needs of specific communities: online voting, juries, petition, elected boards \cite{chandrasekharan_internets_2018} (Irani \& Silberman, 2013; Jhaver et al., 2019; Matias \& Mou, 2018). Contributions include general online governance specification Modular Politics (Schneider et al., 2020), the PolicyKit policy platform \cite{zhang_policykit_2020}, the Digital Juries community-driven conflict resolution system (Fan \& Zhang, 2020), authoring tools that center community (Schneider, 2020) \cite{wang2024pika}, and community-focused analytics platforms \cite{muller2015heapcraft}. Distributed-ledger technologies, particularly surrounding the concept of decentralized autonomous organizations (DAOs) are also inspiring new efforts (Reijers et al., 2018; Wright \& De Filippi, 2015). Still, the contributions in this tradition remain mostly research products. Online communities continue to lack widely available production-scale tools for stewarding their transitions to democratic governance.

\section{Methods}
Governance-focused scholarship has increasingly advocated for explicit textual representations of governance in infrastructure projects, including the recommendation to treat governance documentation as a primary, longitudinal artifact. This trajectory converges with the notion of governance becoming textually instantiated in repository files, offering a concrete path to study how roles, responsibilities, and decision-making rules are defined, communicated, and evolve over time \citep{10.1145/1061874.1061885}\citep{10.1145/3570635}.

After providing the pipeline of the study end-to-end , we describe the corpus, selection criteria, preprocessing, institutional parsing, and analysis that convert governance prose into comparable structures. 

\begin{figure}[t]
  \centering
  \includegraphics[width=\columnwidth,keepaspectratio]{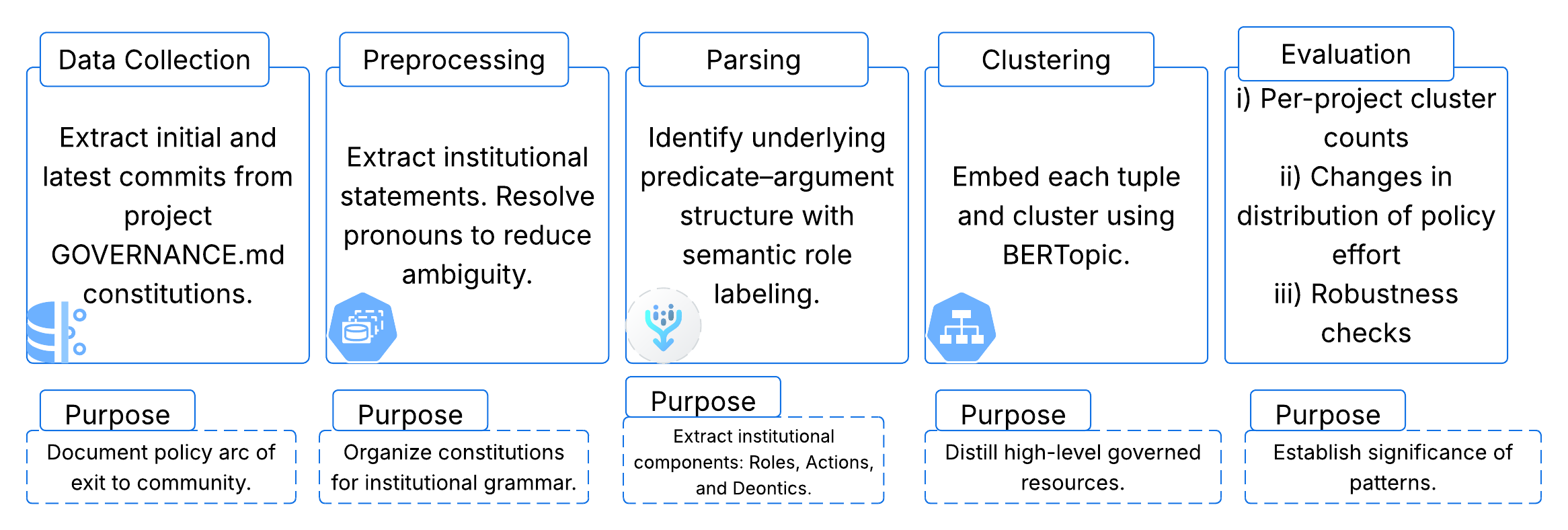}
  \caption{Processing pipeline from raw governance files to structured institutional statements and analysis. These steps support measurement of change in count and concentration from \textit{initial} to \textit{latest} versions of version-controlled \texttt{GOVERNANCE.md} project constitutions. This diagram shows how governance text is normalized, parsed into roles, actions, and deontics, and clustered into institutional constructs. }
  \Description{A left-to-right block diagram showing ingestion of \texttt{GOVERNANCE.md} files, parsing into roles/actions/deontics, clustering, and metrics (count and entropy) feeding result plots.}
  \label{fig:pipeline}
\end{figure}

Figure ~\ref{fig:pipeline} shows our pipeline  transforming raw governance documents into structured institutional data. It begins with data normalization and pairing rules that align governance snapshots across versions. Coreference resolution reduces pronoun ambiguity, enabling more accurate attribution of roles. Semantic Role Labeling (SRL) maps sentences to predicate-argument structures, identifying the underlying grammar of governance actions. These structured tuples are then embedded and clustered using BERTopic to capture governance topologies. The resulting clusters are evaluated using metrics such as entropy and per-project cluster counts to quantify structural diversity, prescriptiveness, and change. This modular pipeline supports scalable, interpretable analysis of institutional evolution in OSS projects.

\textbf{Data and coverage.} 
OSS communities address their governance challenges through informal norms, foundation-level oversight, and increasingly, explicit written constitutions. 
On GitHub, prominent projects are increasingly storing their emergent governance documents in repositories alongside project code, allowing them to benefit from the features of the git software, namely version control for storage of past drafts and easier sharing across projects (“forking”). 
A notable development has been the emergence of \texttt{GOVERNANCE.md} as a de facto standard for codifying project rules, alongside related artifacts such as \texttt{CONTRIBUTING.md}, codes of conduct, and maintainership guides. These files articulate roles, permissions, obligations, and protected resources, making governance unusually transparent and traceable. 

Starting with a seeded collection and filename patterns, we analyzed 710 repositories with at least one governance file at the repository root. The corpus spans 2013–2022, with governance commits recorded through June 2022 (earliest: 2013-05-09; latest: 2022-05-19). File coverage is dominated by \texttt{GOVERNANCE.md}, which appears in 673 out of 710 projects (94.8\%), alongside 37 filename variants. The latest governance file is present for all projects, and Markdown structure is detectable in 498 repositories (70.1\%), with a median of 5 sections (range: 2 to 25) \cite{yan2023github}.

Across the corpus we record 3,889 governance commits corresponding to repository by commit pairs and 2,890 unique commit OIDs, covering 107,869 line level edits with 82,076 additions and 25,793 deletions.

\textbf{Pairing \textit{initial} and \textit{latest} constitutions.} The pipeline produced net governance changes (earliest and latest snapshots) for the 637 repositories over an observation period from 2014-03-26 to 2022-05-18. \texttt{GOVERNANCE.md} file names were dominant, being 601 of 637 (or 94.3\%). Inclusion requires at least two recoverable governance snapshots per project; we label the earliest valid snapshot as \emph{initial} and the most recent as \emph{latest}. We require  at least two distinct \texttt{GOVERNANCE.md} commits falling on different calendar days. For each repository with a governance file, we traversed the Git history to recover the earliest valid version of that file and paired it with the most recent version. Projects with only one usable snapshot we excluded from longitudinal analysis but retained for descriptive statistics. For the 637 paired repositories, the gap from earliest to latest commit has median 0 days with interquartile range 247, minimum 0, and maximum 2616; we refer to 0 day gaps as within day revisions. The across day change subset comprises 279 of 637 which equals 43.8 percent. Where multiple governance files existed, we would create a composite governance view by concatenating in a deterministic order and removing repeated boilerplate; in this paired cohort, one governance file per repository sufficed.

\textbf{Normalization and alignment.} We preprocess governance documents by removing badges and images, converting tables to lists, normalizing headings, and stripping markup. Text we segmented into short paragraph blocks and sentences using a splitter tuned for Markdown lists. To reduce pronoun ambiguity, we applied coreference resolution while maintaining a reversible mapping to original offsets \citep{lee-etal-2018-higher} \citep{jurafskyspeech}. Where headings were detectable, we recorded section counts across snapshots to capture the degree of governance structuring.

\textbf{Institutional parsing.} 
The governance structure of each GitHub project in our corpus we extracted from its \texttt{GOVERNANCE.md} constitution using the Institutional Grammar (IG) framework \citep{ostrom2009understanding, crawford_grammar_1995}. This framework maps the syntactic elements of policy texts to institutional primitives, first decomposing paragraphs and multi-phrase sentences into simple "institutional statements". Under the institutional grammar, an institution is treated as a "bag" or unordered collection of institutional statements. Recent NLP methods have made their automated extraction from policy text feasible~\cite{rice2021machine,chakraborti2024nlp4gov}.

Governance documents were parsed into institutional statements consisting of four linked components. An institutional statement has a \emph{Role} (known in IG as 'Attribute') when its grammatical subject is a kind of agent. Roles account for the types of actor or position recognized by the institution (e.g., "project lead," "contributor"). An \emph{Action} (the `Aim' in IG; syntactically the verb) identifies activities recognized by the institution as requiring governance (e.g., "commit," "assign," "review"). A \emph{Deontic} captures the prescriptive force of the institutional statement, expressed through modal verbs such as "may," "should," or "must," which indicate whether an action is permitted, recommended, or required. Deontics can also be enabling ("can") or restricting ("cannot"). There IG has many other components we do not endeavor to extract, like the \emph{Object}, representing the grammatical object of the rule, whether another actor or a resource that is subject to the action enacted by the statement's role. For example, in the sentence "The technical committee must ratify the development roadmap", the Role is "technical committee," the Action is "ratify," the Object is "roadmap," and the Deontic is "must," which renders the statement obligatory. 

\begin{table}[t]
\centering
\footnotesize
\setlength{\tabcolsep}{10pt}
\renewcommand{\arraystretch}{1.35}
\caption{\textbf{Project roles defined in \texttt{GOVERNANCE.md} files by projects undergoing transition.} We analyze the roles defined by our corpus of \texttt{GOVERNANCE.md} in version control of major OSS projects on GitHub. We do so by extracting 20 kinds of roles defined by constitutions in our corpus, defined here. Key measures in this work around the count of categories that a given constitution invokes, and the entropy of each project's distribution over these types, a measure of how balanced constitutions are in the extent to which they formalize each role. }
\begin{tabularx}{\linewidth}{l >{\raggedright\arraybackslash}X}
\toprule
\textbf{Role Category} & \textbf{Description}  \\
\midrule
all\_project & Rules for all project contributors \\
contributors & Rules for code contributors \\
maintainers & Rules for project maintainers \\
all\_community & Rules for all project stakeholders, including users \\
core\_team & Rules for the closed group of core contributors \\
technical\_committee & Rules for the operation of technical committees \\
subcommittee & Rules for subcommittees of committees \\
the\_project & Rules about how the project itself operates \\
ecosystem & Rules for interfacing with other related projects, such as dependencies \\
oversight & Rules for oversight committee roles \\
meeting\_makers & Rules for the conduct of meetings \\
steering & Rules for the conduct of steering committees \\
misc & Rules for other project roles \\
outside & Rules for interfacing with outside entities and authorities, including governments \\
candidate & Rules for individuals seeking formal positions \\
project\_lead & Rules for the project lead, who may or may not be the project founder \\
reviewers & Rules for those reviewing code \\
chairs & Rules for chairs of committees \\
respected\_members & Rules for respected community members without references to a formal position \\
github & Rules for use of the platform \\
\bottomrule
\end{tabularx}

\label{tab:roles-defined}
\end{table}

\begin{table}[t]
\centering
\footnotesize
\setlength{\tabcolsep}{5pt}
\renewcommand{\arraystretch}{1.35}
\caption{\textbf{Project action categories defined in GOV.md files.} The seven types of action are from the the Institutional Analysis and Design framework for understanding self-governing resource management systems.  We aggregate contextualized verbs observed in the \texttt{GOVERNANCE.md} corpus into these seven categories, to measure the count of categories invoked by project constitutions, and the balance (entropy) of category distributions.}
\begin{tabular}{lll}
\toprule
\textbf{Action Category} & \textbf{Description} & \textbf{Examples}\\
\midrule
Aggregation & For when verbs aggregate individual behaviors into group output & \textit{ discuss, negotiate, decide as a group, contribute}  \\
Position & For when verbs define roles or assign to positions  & \textit{ appoint, compose, serve}  \\
Information & For when verbs describe information operations &  \textit{ document, receive, monitor,  inspect}   \\
Choice & For when verbs describes actor-driven actions or procedural steps & \textit{ submit, build, use, merge, resign}  \\
Constitutive & For when verbs define an institutional entity/abstraction &  \textit{ comprise, include, describe, exist}  \\
Authority & For when verbs assign rights over others & \textit{ approve, assign, require, select, mandate, amend}\\
Payoff & For when verbs invoke exchanges of value & \textit{ pay, distribute, compensate, deposit, penalize}\\
\bottomrule
\end{tabular}
\label{tab:actions-defined}
\end{table}

We extracted these components with the NLP4Gov toolkit \citep{chakraborti2024nlp4gov} \citep{chakraborti2024we}, which combines dependency parsing with semantic role labeling to parse each unitary institutional statement into its IG components. The parser emits tuples with  anchors spans and positions, enabling traceability back to the original text. Modal verbs such as \emph{may}, \emph{can}, \emph{should}, \emph{must}, and \emph{will} were canonicalized into a closed set of deontic types, while role names such as \emph{maintainer}, \emph{committer}, \emph{reviewer}, and \emph{release manager} we normalized manually into a controlled vocabulary. We further manually categorized Actions into a version of the Typology of Rules adapted from the institutional analysis literature~\cite{ostrom2009understanding, weible2012understanding, weible2017policy, weible2018understanding}. To test the reliability of these qualitative steps of the analysis, two authors coded the same sample of 50 Actions, which among the four types of institutional features demanded the most manual categorization. Over this sample they demonstrated a percent agreement of $82\%$ and a Cohen's $\kappa=0.92$, over 9 labels, including a null label, strong evidence for the intersubjective validity of the chosen typology.

\textbf{Embedding, clustering, and metrics.}
Each canonical tuple is rendered as a short governance statement.
We encode each governance statement with a Sentence-BERT encoder \citep{reimers2019sentence} and apply \textsc{BERTopic} \citep{grootendorst2022bertopic} to derive semantic clusters per repository at the \textit{initial} and \textit{latest} snapshots.
\textsc{BERTopic} operates in the embedding space to form topic–like groups and uses class–based TF–IDF to label them. To ensure even clustering across all the projects' corpus, we use the library’s standard hyperparameters without custom tuning.
For structure, we report (i) count $K$ as the number of distinct clusters per repository and (ii) normalized Shannon entropy $H$ (bits, base~2) over cluster proportions, with longitudinal change $\Delta H = H_{\text{latest}} - H_{\text{initial}}$.
For a robustness check, we compute Jensen–Shannon divergence (JSD, bits) between the aligned initial and latest cluster distributions for each repository. 
JSD measures the presence of semantic drift between the initial and latest snapshots by quantifying the dissimilarity between their category distributions. JSD and its generalizations have been used successfully elsewhere as a robust, symmetric drift metric in textual governance analysis \citep{10.48550/arxiv.0804.1653}\citep{10.14736/kyb-2021-6-0879}\citep{10.3390/e21050485}.
All repository–level estimates are aggregated with equal–weight bootstrap confidence intervals by resampling repositories with replacement.

\textbf{Analysis and inference.}
Using the paired across day subset defined above, we compute, for each repository $r$ and snapshot $v\in\{\text{initial},\text{latest}\}$, the empirical distribution over semantic cluster labels. Below we provide the main equations of the methodology. Specifically, normalized Shannon entropy $H_v(r)$ summarizes evenness (Eq.~\ref{eq:entropy}), and \emph{change} is $\Delta H(r)$ (Eq.~\ref{eq:deltaH}).
Distributional change we measured with the Jensen--Shannon divergence between the aligned initial and latest distributions (Eq.~\ref{eq:jsd}).
Count $K_v(r)$ is the number of distinct labels in snapshot $v$ with a presence threshold of at least two statements (Eq.~\ref{eq:richness}); the paired change is $\Delta K(r)$ (Eq.~\ref{eq:deltaK}).
To control for document length, we also report a rarefied $\Delta K$ by sampling an equal number of statements from both snapshots and averaging paired differences over repeated draws (Eq.~\ref{eq:rarefiedDK}).
Entropy and JSD are computed only for repositories with at least five labeled statements in each snapshot; count uses the presence threshold described above.
All repository–level estimands are reported as equal-weight means across repositories with percentile confidence intervals obtained by a repository bootstrap (Eqs.~\ref{eq:bootReplicate}–\ref{eq:bootCI}; resampling repositories with replacement, $B{=}10{,}000$).
Unless noted otherwise, intervals are $95\%$ and units are bits for $H$ and JSD.

Using the paired across-day subset defined above, we compute, for each repository $r$ and snapshot $v\in\{\text{initial},\text{latest}\}$, the empirical distribution over semantic cluster labels. 

Specifically, Normalized Shannon entropy $H_v(r)$ summarizes evenness:
\begin{equation}
H_v(r) \;=\; - \sum_{k} p_{r,v}(k)\,\log_2 p_{r,v}(k),
\quad v\in\{\text{initial},\text{latest}\}.
\label{eq:entropy}
\end{equation}

The \emph{change} is $\Delta H(r)$:
\begin{equation}
\Delta H(r) \;=\; H_{\text{latest}}(r) \;-\; H_{\text{initial}}(r).
\label{eq:deltaH}
\end{equation}

Distributional change is measured with Jensen--Shannon divergence in bits between the aligned initial and latest distributions, itself a difference in Kullback-Leibler divergences:
\begin{equation}
\mathrm{JSD}\!\left(p_{r,\text{initial}},p_{r,\text{latest}}\right)
\;=\; \tfrac{1}{2}\,\mathrm{KL}\!\left(p_{r,\text{initial}} \,\middle\|\, m_r\right)
\;+\; \tfrac{1}{2}\,\mathrm{KL}\!\left(p_{r,\text{latest}} \,\middle\|\, m_r\right),
\quad m_r=\tfrac{1}{2}\!\left(p_{r,\text{initial}}+p_{r,\text{latest}}\right).
\label{eq:jsd}
\end{equation}

Count $K_v(r)$ is the number of distinct labels in snapshot $v$ with a presence threshold of at least two statements:
\begin{equation}
K_v(r) \;=\; \sum_{k}\mathbf{1}\!\left\{\,c_{r,v}(k)\ge \tau\,\right\}, 
\qquad \tau=2.
\label{eq:richness}
\end{equation}

The paired change is $\Delta K(r)$:
\begin{equation}
\Delta K(r) \;=\; K_{\text{latest}}(r) \;-\; K_{\text{initial}}(r).
\label{eq:deltaK}
\end{equation}

To control for document length, we also report a rarefied $\Delta K$ by sampling an equal number of statements from both snapshots and averaging paired differences over repeated draws:
\begin{equation}
\widetilde{\Delta K}(r) \;=\; \frac{1}{R}\sum_{t=1}^{R}
\Big( K^{(t)}_{\text{latest}}(r;n_r) \;-\; K^{(t)}_{\text{initial}}(r;n_r) \Big),
\quad n_r=\min\!\big\{N_{r,\text{initial}},\,N_{r,\text{latest}},\,100\big\}.
\label{eq:rarefiedDK}
\end{equation}

Entropy and JSD are computed only for repositories with at least five labeled statements in each snapshot; count $K$ uses the presence threshold described above. All repository–level estimands are reported as equal-weight means across repositories with percentile confidence intervals obtained by a repository bootstrap (resampling repositories with replacement, $B{=}10{,}000$). For each bootstrap replicate:
\begin{equation}
\hat{\theta}^{*(b)} \;=\; \frac{1}{n}\sum_{r \in \mathcal{R}^{*(b)}} s(r),
\qquad b=1,\dots,B,\; B=10{,}000.
\label{eq:bootReplicate}
\end{equation}

Confidence intervals are defined as:
\begin{equation}
\mathrm{CI}_{1-\alpha} \;=\;
\Big[\, Q_{\alpha/2}\big(\{\hat{\theta}^{*(b)}\}\big),\;
Q_{1-\alpha/2}\big(\{\hat{\theta}^{*(b)}\}\big) \,\Big].
\label{eq:bootCI}
\end{equation}

Unless noted otherwise, intervals are $95\%$ and units are bits for $H$ and JSD.

\begin{table}[t]
\centering
\footnotesize
\setlength{\tabcolsep}{15pt}
\renewcommand{\arraystretch}{1.15}
\caption{\textbf{Projects define more roles and govern more actions over time.}
Table reports the values of $K$, the average number of roles, action, or deontic categories used in a \texttt{GOV.md} constitution, and the difference between \textit{initial} and \textit{latest}. Deontic verbs (from "can" and "should" to "must") capture the average prescriptiveness of an institutional statement. Bootstrapped 95\% CIs are over repositories (\(B{=}10{,}000\)). The rarefied estimate draws the same
number of statements from each snapshot (cap 100) before counting, to control for
later constitutions being longer. Units are counts of distinct clusters; bold intervals exclude 0.}
\label{tab:count-results}
\begin{tabular}{lccccc}
\toprule
\textbf{Feature} & \textbf{$n$} &
\textbf{Initial $K$} & \textbf{Latest $K$} &
\textbf{Mean $\Delta K$ [95\% CI]} & \textbf{Rarefied $\Delta K$ [95\% CI]} \\
\midrule
Roles    & 244 & 3.46 & 3.95 & \textbf{+0.484 [0.258, 0.713]} & \textbf{+0.224 [0.092, 0.352]} \\
Actions  & 266 & 3.86 & 4.46 & \textbf{+0.602 [0.417, 0.793]} & \textbf{+0.228 [0.134, 0.326]} \\
Deontics & 236 & 1.14 & 1.15 & +0.008 [$-0.038$, 0.055]        & $-0.024$ [$-0.062$, 0.012]      \\
\bottomrule
\end{tabular}
\end{table}
\begin{table}[t]
\centering
\footnotesize
\setlength{\tabcolsep}{15pt}
\renewcommand{\arraystretch}{1.15}
\caption{\textbf{Attention to roles and actions becomes more balanced across categories while polarity of deontics ("can" versus "cannot") becomes more concentrated on enabling verbs.}
Table reports changes in concentration (Shannon entropy \(\Delta H = H_{\text{latest}}-H_{\text{initial}}\), bits) (mean) and within–repository distributional change (Jensen–Shannon divergence, bits). Deontic$^{\dagger}$ shows results for binary coding of deontic verbs into enabling vs.\ restricting. Rows show means across repositories; 95\% CIs are from equal–weight bootstrapping over repositories (\(B{=}10{,}000\)). Bold \(\Delta H\) intervals exclude 0. }
\begin{tabular}{lcccccc}
\toprule
\textbf{Feature} & \textbf{$n$} & \textbf{Initial $H$} & \textbf{Latest $H$} & \textbf{$\Delta H$ [95\% CI]} & \textbf{JSD [95\% CI]} \\
\midrule
Roles   & 169 & 1.775 & 1.866 & \textbf{+0.092 [0.011, 0.173]} & \textbf{0.202 [0.172, 0.234]} \\
Actions & 213 & 1.905 & 1.979 & \textbf{+0.074 [0.017, 0.134]} & \textbf{0.126 [0.107, 0.146]} \\
Deontic & 144 & 1.057 & 1.052 & $-0.005$ [$-0.066$, 0.056]     & \textbf{0.062 [0.048, 0.079]} \\
Deontic$^{\dagger}$ & 149 & 0.108 & 0.076 & $\textbf{-0.032}$ [$\textbf{-0.066}$, $\textbf{-0.001}$] & \textbf{0.009 [0.005, 0.014]} \\
\bottomrule
\end{tabular}
\label{tab:entropy-results}
\end{table}


\section{Results}

In this paper, we analyze the change within a repository by pairing the earliest recoverable \texttt{GOVERNANCE.md} snapshot with the latest and computing: (i) Shannon entropy for each version \(H_v(r)\) in Eq.~\eqref{eq:entropy} and the paired change \(\Delta H(r)\) in Eq.~\eqref{eq:deltaH}; (ii) the count of distinct constructs \(K_v(r)\) in Eq.~\eqref{eq:richness} and its paired change \(\Delta K(r)\) in Eq.~\eqref{eq:deltaK}, with the size controlled rarefied estimate \(\widetilde{\Delta K}(r)\) in Eq.~\eqref{eq:rarefiedDK}. We also use Jensen–Shannon divergence to measure the change in construct distributions between snapshots (Eq.~\eqref{eq:jsd}). We aggregate repository-level summaries with equal-weight bootstrap confidence intervals using the resampling scheme in Eqs.~\eqref{eq:bootReplicate}–\eqref{eq:bootCI} \((B{=}10{,}000)\). Unless noted, units are bits for \(H\) and JSD. A minimum of five labeled statements is required per version screen, as described in Methods.

\begin{figure}[!ht]
  \centering
  \includegraphics[width=0.7\columnwidth]{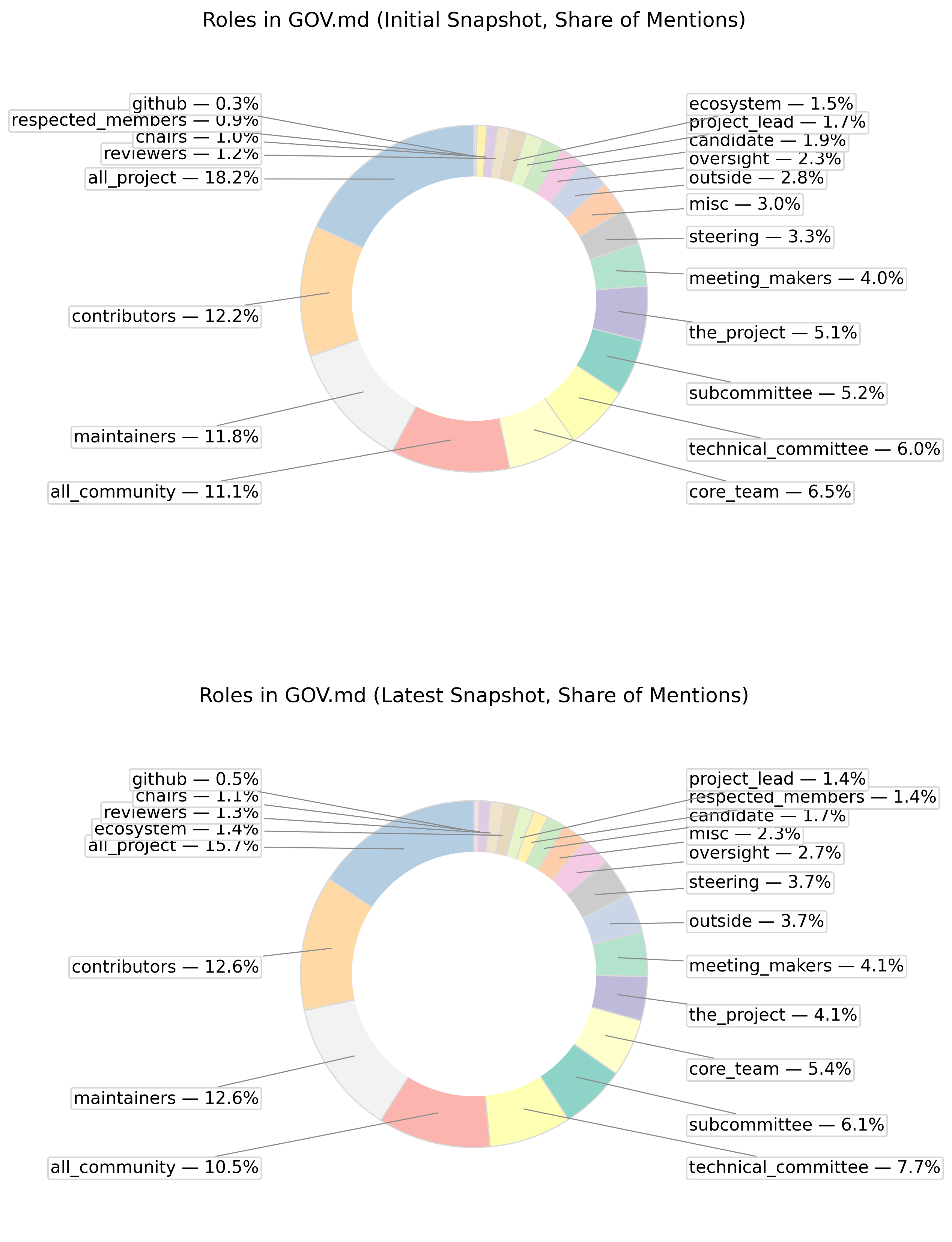}
    \caption{\textbf{Projects diversify the range of governance roles over time.} Plots show the share of role mentions in initial versus latest governance snapshots. Early constitutions are dominated by broad categories such as ``all\_project'' and ``all\_community,'' while later constitutions redistribute attention across more specialized roles (e.g., subcommittees, technical committees, and steering groups). This broadening reflects institutional development toward greater specialization and shared governance. The differences in the distributions of these types are small but significant.}

  \Description{}
  \label{fig:donut_roles}
\end{figure}

\begin{figure}[!ht]
  \centering
  \includegraphics[width=0.7\columnwidth]{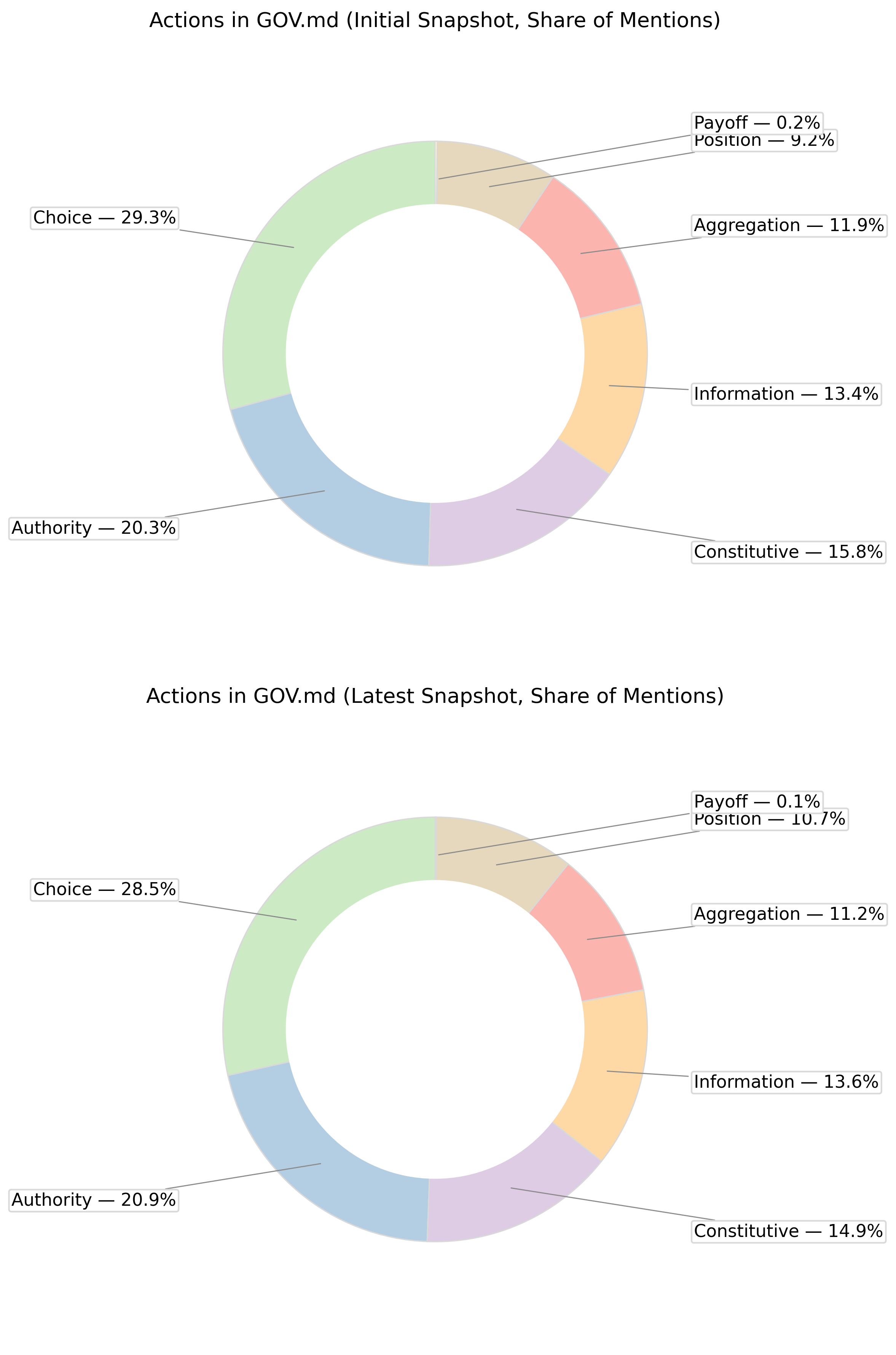}
    \caption{\textbf{Projects expand the catalog of governance actions over time.} Plot shows the share of action mentions in initial versus latest governance snapshots. While high-level categories such as ``choice'' and ``authority'' remain prominent, later constitutions show a broader and more balanced distribution across action types, reflecting increased institutional complexity and scope. As above, the differences in the distributions of these types are small but significant.}
  \Description{}
  \label{fig:donut_actions}
\end{figure}



\begin{figure}[!ht]
  \centering
  \includegraphics[width=1.1\columnwidth]{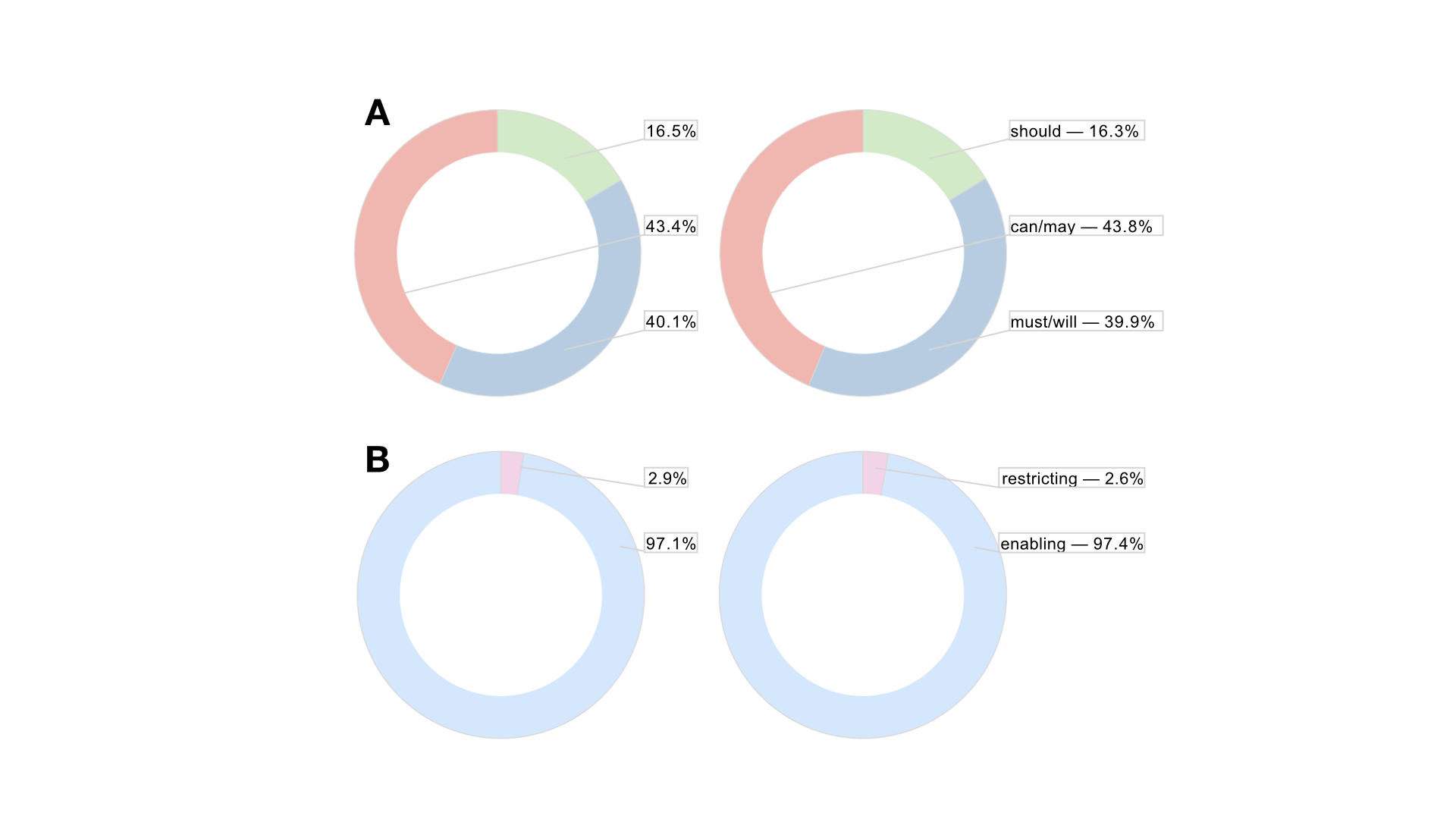}
    \caption{\textbf{Deontic composition in OSS governance remains broadly stable over time, although entropy of binomial enabling/restricting distribution increases.} Panel A show the share of modal expressions (can/may,'' must/will,'' should'') in initial versus latest governance snapshots. The relative balance between permissive, obligatory, and advisory language changes little, suggesting that while projects diversify roles and actions, the prescriptive force of their rules remains largely constant. Panel B highlights the distribution of enabling versus restricting deontic statements. Enabling language (``can'', ``may'') accounts for over 97\% of references in both periods, while restricting terms (``cannot'', ``must not'') remain a small minority, and seem to decline further over time. Together, these patterns underscore the relative stability of governance constitutions, where permissions dominate over prohibitions.}
  \Description{}
  \label{fig:donut_deontic_2}
\end{figure}

\begin{figure}[!ht]
  \centering
  \captionsetup{font=small}
  \begin{subfigure}[t]{0.5\columnwidth}
    \centering
    \includegraphics[width=\linewidth]{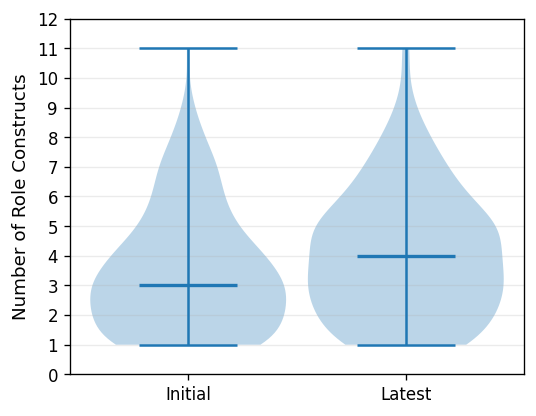}
    \subcaption{Roles: initial vs.\ latest}
    \label{fig:violin-role}
  \end{subfigure}\hfill
  \begin{subfigure}[t]{0.5\columnwidth}
    \centering
    \includegraphics[width=\linewidth]{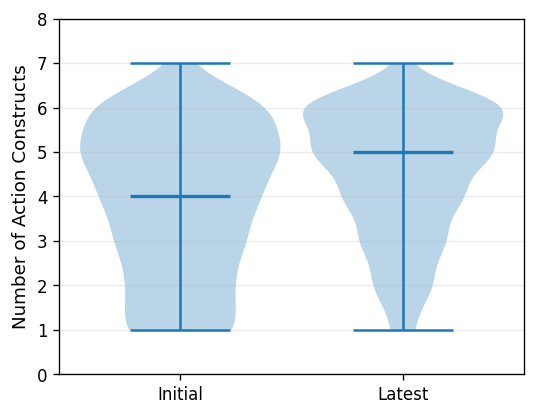}
    \subcaption{Actions: initial vs.\ latest}
    \label{fig:violin-action}
  \end{subfigure}\hfill
  \begin{subfigure}[t]{0.5\columnwidth}
    \centering
    \includegraphics[width=\linewidth]{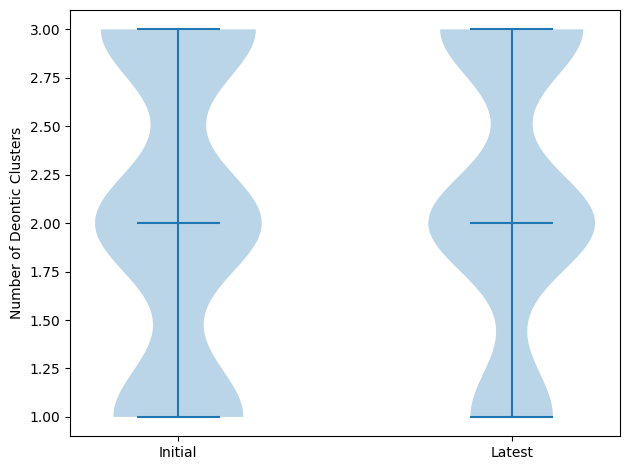}
    \subcaption{Deontics: initial vs.\ latest}
    \label{fig:violin-deontic}
  \end{subfigure}
  \vspace{-2mm}
  \caption{\textbf{The number of types of roles and actions invoked in constitutions increases significantly.} Violin plots compare the number of categories of (a) Roles, (b) Actions, and (c) Deontics between initial and latest snapshots for each repository. Roles and actions both show significant increases.}
  \Description{Three side-by-side violin plots. The first shows role distributions for initial and latest snapshots, the second shows action distributions, and the third shows deontic distributions; each violin reflects per-repository values across the corpus.}
  \label{fig:violin-panel}
  \vspace{-2mm}
\end{figure}

\textbf{Count $\textbf{K}$}.
Projects define a wider array of who acts and what is governed over time. Roles show clear increases in the number of distinct constructs per repository (Fig.~\ref{fig:violin-role}), and actions follow a similar pattern (Fig.~\ref{fig:violin-action}). These increases remain positive under the rarefied control that equalizes snapshot length \(\big(\widetilde{\Delta K}(r)\) in Eq.~\eqref{eq:rarefiedDK}\big). By contrast, deontic counts show no effect on average (Fig.~\ref{fig:violin-deontic}).
Table~\ref{tab:count-results} summarizes the paired change in the number of distinct
role and action clusters between initial and latest governance snapshots.
Both features show positive mean changes, and the corresponding bootstrap intervals
exclude zero, indicating a consistent broadening of who acts and what is governed across repositories. Looking more deeply into the typical distributional changes, we find that the increase in entropy of the distribution over project roles is accounted for in part by increases in the regulation of projects' connections with their ecosystems, and their growing invest in regulations specifying defining oversight roles. Rarefied estimates remain positive, confirming that these findings are
not artifacts of the fact that later constitutions tend to be longer. Together, these results provide strong evidence
that projects expand the scope of their governance texts as they mature, formalizing
a more diverse set of actors and activities.

\textbf{Concentration.}
With time, constitutions distribute proportionately more attention across their more specialized roles (Fig.~\ref{fig:donut_roles}). 

In terms of roles (Fig.~\ref{fig:donut_roles}) and actions (Fig.~\ref{fig:donut_actions}), both constructs become more balanced in how their categories are formalized, with more even distribution of policy over categories with time. Mean $(\Delta H)$ is positive for both features and the corresponding intervals exclude zero. Deontic verbs stay unchanged in their relative prescriptiveness ('may's versus 'should's and 'must's) (Fig.~\ref{fig:donut_deontic_2}), but they do show a change when aggregated along an enabling versus restricting dichotomy ('can' versus 'cannot') (Fig.~\ref{fig:donut_deontic_2}), demonstrating a relative decrease in proscriptions with time, despite the general growth in the size and complexity of projects over time. The more simple and striking finding is that enabling deontics outnumber restricting deontics at a rate of $40:1$. This is certainly due in part to the inherently volunteer nature of OSS contributions, but could still be considered extreme. 

Table~\ref{tab:entropy-results} shows that roles and actions have positive $\Delta H$,
indicating a more even distribution of formal policy across each  construct's categories. Represented in terms of prescriptiveness, deontics remain stable, while they increase in concentration ($-\Delta H$) when represented in terms of restrictiveness, showing a reduction in restricting conditions over time. 

Across roles, actions, and deontics, values of Jensen-Shannon divergence, measuring the extent of distributional change between initial and latest snapshots, correspond to those suggested by our other measures. This supporting finding bolsters the internal validity and consistency of our claims of structural diversification over time.

\noindent\textbf{Roles and actions.}
Table~\ref{tab:actions-slice-deltas-rarefied} reveals that action attention shifts away from
\textbf{choice} ($-6.1$ pp) and \textbf{authority} ($-3.9$ pp), with additional declines in
\textbf{aggregation} ($-3.2$ pp) and \textbf{constitutive} actions ($-2.5$ pp), suggesting a move
away from concentrated decision rights and unilateral control toward a more distributed mix of
actions. Table~\ref{tab:roles-slice-deltas-rarefied} shows a similar pattern for roles: broad
categories like \textbf{all\_project} ($-3.8$ pp) and \textbf{the\_project} ($-2.2$ pp) decline,
while attention redistributes across more specialized categories such as \textit{ecosystem}
(+1.9 pp) and \textit{chairs} (+0.2 pp). These results indicate that projects mature by layering
and diversifying responsibilities, broadening governance participation rather than simply
shifting control from one group to another.

\textbf{Interpretation of intervals.}
Comparisons between initial and latest constitutions are bootstrapped on a within-project basis, improving their precision. Our robustness checks control for potential artifacts due to some projects having very short constitutions or very short intervals between initial and latest versions. Another robustness check, our "rarefied" resampling, improves comparisons between initial and latest by controlling for the fact that 'latest' constitutions are almost always longer. We chose not to report our findings from a fourth construct of the institutional grammar, "Objects", because of a mix of difficulties extracting, categorizing, and interpreting them.

\begin{table}[t]
\centering
\footnotesize
\setlength{\tabcolsep}{20pt}
\renewcommand{\arraystretch}{1.35}
\caption{\textbf{The proportions of the seven categories of action stay stable across snapshots. } Entries report repository–paired mean $\Delta$share = share$_{\text{latest}}$ - share$_{\text{initial}}$ with 95\% bootstrap CIs over repositories ($B{=}10{,}000$). Despite the reliable increase in counts, none of the proportions change measurable. Proportion columns sum to 100\%, while difference column sums to 0. Values are percentage points (pp).}
\begin{tabular}{lrrr}
\toprule
\textbf{Action Category} & \textbf{Initial (\%)} & \textbf{Latest (\%)} & \textbf{$\Delta$share (pp) [95\% CI]} \\
\midrule
Aggregation & 12.88 & 11.78 & -1.10 [-3.34, +0.99] \\
Position & 8.80 & 9.85 & +1.05 [-0.85, +2.78] \\
Information & 13.39 & 14.44 & +1.05 [-1.25, +3.26] \\
Choice & 29.59 & 28.96 & -0.63 [-3.17, +1.93] \\
Constitutive & 15.72 & 15.54 & -0.18 [-2.35, +1.95] \\
Authority & 19.55 & 19.38 & -0.17 [-2.35, +1.93] \\
Payoff & 0.07 & 0.05 & -0.03 [-0.09, +0.03] \\
\bottomrule
\end{tabular}
\label{tab:actions-slice-deltas-rarefied}
\end{table}

\begin{table}[t]
\centering
\footnotesize
\setlength{\tabcolsep}{20pt}
\renewcommand{\arraystretch}{1.35}
\caption{\textbf{Roles defining ecosystem, oversight committees, and the project itself show significant changes between snapshots.} Entries report repository–paired mean $\Delta$share = share$_{\text{latest}}$ - share$_{\text{initial}}$ with 95\% bootstrap CIs over repositories ($B{=}10{,}000$). Values are percentage points (pp). Proportion columns sum to 100\%, while difference column sums to 0. Bold rows indicate CIs that exclude 0. }
\begin{tabular}{lrrr}
\toprule
\textbf{Role Category} & \textbf{Initial (\%)} & \textbf{Latest (\%)} & \textbf{$\Delta$share (pp) [95\% CI]} \\
\midrule
all\_project & 17.67 & 15.83 & -1.84 [-4.49, 0.68] \\
\textbf{the\_project} & \textbf{5.52} & \textbf{3.81} & \textbf{-1.70} \textbf{[-3.43, -0.11]} \\
outside & 2.68 & 4.36 & +1.68 [-0.01, +3.42] \\
\textbf{oversight} & \textbf{1.55} & \textbf{2.70} & \textbf{+1.15} \textbf{[+0.04, +2.59]} \\
meeting\_makers & 4.99 & 4.13 & -0.86 [-2.63, 0.77] \\
ecosystem & 1.63 & 2.44 & +0.81 [-0.02, +1.95] \\
contributors & 11.40 & 12.13 & +0.74 [-1.94, +3.43] \\
core\_team & 5.44 & 4.80 & -0.65 [-1.78, 0.48] \\
all\_community & 9.55 & 10.15 & +0.59 [-1.33, +2.54] \\
misc & 2.52 & 2.06 & -0.46 [-1.54, 0.52] \\
project\_lead & 2.20 & 1.74 & -0.45 [-1.51, 0.47] \\
subcommittee & 5.24 & 5.65 & +0.40 [-1.34, +2.00] \\
chairs & 0.50 & 0.84 & +0.34 [-0.04, +0.80] \\
maintainers & 16.78 & 16.49 & -0.28 [-3.08, 2.45] \\
github & 0.27 & 0.51 & +0.23 [-0.14, +0.75] \\
respected\_members & 0.83 & 1.01 & +0.18 [-0.16, +0.56] \\
technical\_committee & 6.46 & 6.62 & +0.16 [-1.54, +1.80] \\
reviewers & 0.82 & 0.93 & +0.11 [-0.39, +0.64] \\
steering & 2.64 & 2.55 & -0.09 [-1.21, 0.97] \\
candidate & 1.32 & 1.26 & -0.06 [-0.60, 0.51] \\
\bottomrule
\end{tabular}
\label{tab:roles-slice-deltas-rarefied}
\end{table}


\section{Discussion}

In their transitions from unitary to community governance, communities grow, diversify, and rebalance their regulation of their core actors and activities. These patterns show that projects implement community management by first narrowly laying out key roles and actions, and then expanding and distributing formal regulation of them with time. 

Although our design is presented as a comparison of two snapshots, it is in reality a comparison of three, the "default" \textit{empty} (undefined, 0-byte) \texttt{GOVERNANCE.md} file that the project had before initiating it's transition from GitHub's software default of unitary project management, the \textit{initial} \texttt{GOVERNANCE.md} file representing the community's earlier formal expressions of its community management, and the \textit{latest} presenting the most mature version of its constitution to-date. Comparing across these three states, we find evidence that, as communities grow their constitutions, they begin by defining the key roles and actions that they expect to use. As they mature, they create more but also flesh out what they have, as evidenced by the increased in both the counts of roles and actions and the entropies of their distributions (Tables~\ref{tab:count-results} and \ref{tab:entropy-results}). 

These patterns suggest that as communities mature, governance takes on an increasingly coordinating role: more of the governance “work” is about defining, distributing, and orchestrating roles and responsibilities across the project. This is consistent with the voluntary nature of open source—where governance must facilitate collaboration rather than simply control it. 

Descriptively, roles show a greater concentration of policy in broad categories such as \textbf{all\_project} and \textbf{all\_community} remain among the largest slices, consistent with an approach focused on defining general obligations for all stakeholders (boundary rules) in parallel with more narrow role-specific responsibilities.  Contributors and maintainers remain among the largest groups ($\approx25\%$), which is intuitive given they are the productive core of projects. These results give practitioners a concrete signal that investing in clear definitions of contributor roles and review mechanisms is not just bureaucratic overhead but a sign of project health and maturation as communities become secure in their self-governance.

The specific category changes constituting the distributional change are increases in regulation of \textbf{oversight} roles. These changes suggest that community governance increasingly formalizes relationships with external stakeholders and institutes decision review processes. The increases are balanced by decreases in the proportion of regulation for many other roles, all insignificant except the decrease in references to \textbf{the\_project}. Policy statements about the project in the abstract decrease over time in favor of statements identifying specific individuals, bodies, and populations, consistent with a recognition of the importance of well-defined roles and the explicit assignment of responsible parties to all of a project's core functions.

Some of the most instructive findings are the null effects. The stability of \textbf{project\_lead} roles indicates that projects neither consolidate nor dismantle formal leadership positions over time. Roles of leadership coexist with broader participation. For those managing or founding OSS projects, this implies that early governance documents carry long-term weight: decisions made in the first committed constitution tend to persist and shape all subsequent evolution.

Among actions, the largest category remains \textbf{choice}, prescribing individual action, followed by \textbf{constitutive} and \textbf{authority} type actions which are concerned with defining institutional constructs and assigning their management. The persistence of authority signals that transitions are not about eliminating power but redistributing it among stakeholders. Other major categories—\textbf{information} (which specifies information flows) and \textbf{aggregation} (the manner in which individual behaviors get organized into the collective's behavior) reinforce the view of governance as scaffolding for collective work. 

According to our results, all of these types of action remain in consistent proportion with respect to each other. This stability of action categories, if it has implications for institutional structure and design, suggestions that the types of actions are equally relevant at all stages of the governance transition process. In particular, the limited change in \textbf{aggregative} and \textbf{constitutive} actions is consistent with the theory that the fundamental constitutional structure of projects is set early and that most subsequent development refines rather than rewrites it.

\subsection{Contributions}

These observations have broader implications for the CHI community and for public-interest technology. First, they show that governance infrastructure in open source functions less as a static rulebook and more as a coordination mechanism that evolves as participation grows. Designing tools that make governance text easier to draft, revise, and compare across projects could help communities adopt good practices earlier. Second, the finding that authority persists but becomes more distributed invites platform designers to build affordances for transparent decision-making, not just for formal voting but also for documenting rationale, oversight, and accountability structures. Finally, the layered nature of change—growth through accretion rather than replacement—suggests that interventions that lower the cost of adding clarifying roles, committees, or procedures may support more inclusive governance without requiring disruptive constitutional rewrites.

Entropy gains and rarefied $\Delta K$ results confirm that diversification is not simply an artifact of the fact that later constitutions are longer. Jensen–Shannon divergence indicates meaningful within-project drift. This has practical implications for researchers and toolmakers: monitoring change points in governance text may serve as an early indicator of institutional transition moments, which could be surfaced to community members or funders as signals of project evolution.

Our analysis focuses on governance text rather than behavior, intervals reflect between-project uncertainty via equal-weight resampling, and most artifacts are in English with policies sometimes distributed across multiple files. Our pipeline is artifact-agnostic and can extend to multi-file policy inventories. Equal weighting avoids letting large projects dominate and aligns with participatory aims. We acknowledge that with roughly twenty slices per feature, some significant results may be false positives by chance; we therefore interpret patterns holistically and emphasize consistent signals across complementary measures.


\section{Limitations}
There are, naturally, many limitations to this work. Foremost, we analyze governance text rather than behavior, making this work vulnerable to the commonplace observation that an institution's rules-in-use and rules-in-form typically differ. Another concern is that many consequential rules may live outside \texttt{GOVERNANCE.md} (for example \texttt{CONTRIBUTING.md}, \texttt{CODEOWNERS}, CI settings, issue templates) or in informal channels. Additionally, the corpus is restricted to the English-literate open source community, which could further limit the generality of the findings.
Methodologically, our paired design requires two recoverable snapshots per repository, so survivorship and timing effects can bias change estimates, and stabilization thresholds (at least five labeled statements per snapshot for balance and JSD, presence threshold $\tau=2$ for count $K$) trade variance for selection.
Moving to the analytic approach of this work, natural language processing and representation choices, including segmentation, coreference, embedding, and clustering, can miss conditionality and shape absolute values of $K$, $H$. Within our analysis, equal-weight repository bootstrapping reflects between project variability without fully adjusting for confounders such as age or scale, while our robustness measure, Jensen Shannon divergence, can be affected by cluster relabeling across snapshots. The qualitative components of the analysis could also have introduced distortion, or limit reproducibility. Although coder agreement for Action types was high, residual category bias is possible. And finally, our inference is ultimately comparative and correlational, rather than causal. 

\section{Conclusion}

Open source software (OSS) underlies critical infrastructure around the globe, and has a hand in every aspect of digital life. And yet, most OSS projects are run as founder-led, single-leader governance model qwith no formal accountability to their community, or society. Fortunately, OSS projects are not waiting for policymakers to step up. Increasing numbers of prominent OSS projects are voluntarily transitioning their governance institutions from their founder-owners to their communities. How do projects structure these transitions? 

We find increases in the number of roles and actions that projects define, as well as the entropies of the distribution of institutional statements over those constructs. These patterns are consistent with maturation of constitutions by accretion: institutional prescriptions grow, broaden, and rebalance, while prescriptive polarity changes slowly. 
With this contribution, we hope that critical OSS projects will face more predictable risks as they navigate a dynamic policy landscape and shifting community needs. By addressing these challenges, we can capture the institutional context of accountable OSS design, and develop general tools for transitioning software projects to accountable governance.

\bibliographystyle{ACM-Reference-Format}
\bibliography{references}

\end{document}